\documentclass[graphics,nofootinbib,tightenlines,aps,pra,showpacs,twocolumn,reprint]{revtex4-1}
\usepackage{graphicx,graphics,epsfig}
\usepackage{dcolumn}
\usepackage{bm}
\usepackage{amsmath}
\usepackage{verbatim}
\usepackage{color}
\usepackage[colorlinks=false]{hyperref} 
\usepackage{subfigure}
\usepackage{times,natbib}
\usepackage{amsmath,amsfonts,amssymb,graphics,graphicx,epsfig,color,times,natbib}
\usepackage{tikz}
\usepackage{pgfplots}
\usepackage{verbatim}

\newcommand{\Exp}{\mbox{exp}}
\begin{document}
\title{Local Unitary Representation of Braids and N-Qubit Entanglements}

\author{Li-Wei Yu}
\email{nkyulw@yahoo.com}
\affiliation{Theoretical Physics Division, Chern Institute of Mathematics, Nankai University,
Tianjin 300071, China }
\affiliation{Department of Mathematics, University of California, Santa Barbara, California 93106, USA}




\begin{abstract}

\textbf{ In this paper, by utilizing the idea of stabilizer codes, we give some relationships between one local unitary representation of braid group in N-qubit tensor space and the corresponding entanglement properties of the N-qubit pure state $|\Psi\rangle$, where the N-qubit state $|\Psi\rangle$ is obtained by applying the braiding operation on the natural basis. Specifically, we show that the separability of  $|\Psi\rangle=\mathcal{B}|0\rangle^{\otimes N}$ is closely related to the diagrammatic version of the braid operator $\mathcal{B}$. This may provide us more insights about the topological entanglement and quantum entanglement.}

\end{abstract}


\maketitle



\section{Introduction}
Braid operators have been widely used for quantum information and computation. Especially in topological quantum computation, the processes of braiding anyons, usually related to the unitary Jones representation of braids, act as the role of unitary quantum gates that are immune to local errors\cite{Kitaev20062,nayak2008non,wang2010topological}.  On the other hand, the local unitary representations of braids\cite{Rowell2012}, which are different from the Jones representation, have been also well connected to the quantum information and quantum computation\cite{kauffman2002quantum,kauffman2004braiding,chen2007braiding,DELANEY2016}. In terms of the relationships between braid and quantum entanglement, one of the simplest examples is that a special $4\times 4$ braid matrix generates Bell basis from the 2-qubit natural basis\cite{kauffman2004braiding}, where Bell basis represents the maximal entangled 2-qubit pure state. This interesting result has made a well connections  between the braid operators and quantum entanglement.  After that, a series of generalized works have been made\cite{zhang2005universal,zhang2005yang,zhang2007ghz,chen2007braiding,ge2012yang}. One significant generalization is that the parametrized form of the braid relation, Yang-Baxter equation, has been used for describing the entangled degree of pure states, such as 2-qubit\cite{zhang2005universal} and 3-qubit\cite{yu2014factorized}. Besides that, other further investigations associated with Yang-Baxter equation and generalized Yang-Baxter equation  are also made to generate specific N-qudit entangled states\cite{rowell2016parameter,rowell2010extraspecial}. Based on the previous progresses in generating multipartite entanglements, we then come up with a natural question: Are there any  general relationships between braid and quantum entanglement?

In this paper, based on the local unitary representation of braid group associated with Ising theory\cite{wang2010topological,Rowell2012}, we discuss the general relationship between the local unitary representation of N-strand braid group in tensor product space $(\mathbb{C}^2)^{\otimes N}$ and the entanglement of N-qubit final state obtained by applying braid operators on the initial tensor product state. Our results show that the entangled  parties of the final state generated by braiding operation  depend only on the permutations of the strands in the diagrammatic version. In other words, only the permutation group, as a quotient group of the braid group, entangles the qubit sites. 

Here we adopt the idea that if the initial state is stabilized by a stabilizer set $\mathcal{S}$, then the final state after braiding operation $\mathcal{B}$ is also stabilized by the final stabilizer set $\mathcal{S}'=\mathcal{BSB^\dag}$.  Bravyi's paper \cite{PhysRevA.73.042313} shows that for the Majorana representation of braid group in Ising anyon theory, the final stabilizer set after braiding is just a permutation of the site number of the Majorana operators in the initial stabilizer set.  Because of the equivalence between our Pauli version of braid representation and the Majorana version, we can utilize the similar properties of the Majorana stabilizer set to our case.

Now we briefly introduce the logic of the proof. Firstly, we have local unitary representation of braids expressed by Pauli matrices and an initial state $|\Psi_0\rangle$ in tensor product space  $(\mathbb{C}^2)^{\otimes N}$. Secondly, we transform the braid representation of Pauli version into the equivalent Majorana version  by using Jordan-Wigner transformation. Then,  we choose stabilizer set $\mathcal{S}$ including N independent stabilizers for the initial state $|\Psi_0\rangle$.  Applying braiding operation $\mathcal{B}$ on state $|\Psi_0\rangle$, the final state $|\Psi\rangle=\mathcal{B}|\Psi_0\rangle$ is also stabilized by the final stabilizer set $\mathcal{S}'=\mathcal{BSB^\dag}$.  At last,  we use the final stabilizer set $\mathcal{S}'$ to classify the entangled parties of the final state $|\Psi\rangle$. In this paper, we call a state entangled if it cannot be separated into any two parties.  The paper is organized as follows. In Sec. \ref{Sec2}, we introduce the local unitary representation of braids in Pauli version and the equivalent  Majorana version. In Sec. \ref{Sec3}, we give some examples and detailed explanations about the braiding operation and qubit entanglement. In Sec. \ref{Sec4}  we discuss a general case of braiding and entanglement including arbitrary number of qubits. In the last section, we make conclusions and discussions.

\section{Braid group Representation in Pauli Version and Majorana Version}\label{Sec2}
In this section, we mainly introduce one local unitary representation of N-strand braid group $B_N$ in N-qubit tensor product space $(\mathbb{C}^2)^{\otimes N}$ and its equivalent Majorana fermionic version.

The N-strand braid group $B_N$ is presented by generators $\{\tau_i|i=1,2,...N-1\}$ with the relations
\begin{eqnarray}
&&\tau_i\tau_j=\tau_j\tau_i \qquad \text{if} \,\, |i-j|\geq2,\\
&&\tau_i\tau_{i+1}\tau_i=\tau_{i+1}\tau_i\tau_{i+1}.
\end{eqnarray}
One local unitary representation of $B_N$ in N-qubit space $(\mathbb{C}^2)^{\otimes N}$ is 
\begin{equation}\label{BraidPauli}
\begin{aligned}
\tau_i&=\frac{1}{\sqrt{2}}\left({\mathbb{I}}^{\otimes N}-\text{i}{\mathbb{I}}^{\otimes i-1}\otimes\sigma_i^y\otimes\sigma_{i+1}^x\otimes{\mathbb{I}}^{\otimes N-i-1} \right),\\
&=\Exp\left[-\text{i}\frac{\pi}{4}\sigma_i^y\otimes\sigma_{i+1}^x \right],
\end{aligned}
\end{equation}
where $\mathbb{I}$ represents 2D identity matrix and $\sigma_i^{x(y)}$ represents the usual Pauli X(Y) matrix on the i-th site. This is the Pauli version of the local unitary braid representation, which is related to Ising theory.

It is easy to verify that by applying $\tau_i$ to the initial tensor product basis $|\Psi_0\rangle=|0\rangle^{\otimes N}$,\footnote{$|0\rangle$ and $|1\rangle$ are two eigenvectors of Pauli Z matrix $\sigma^z$ in $\mathbb{C}^2$, where $\sigma^z|0\rangle=-|0\rangle$, $\sigma^z|1\rangle=|1\rangle$.} one obtains the entangled 2-qubit Bell state on i-th and (i+1)-th sites\cite{kauffman2004braiding},
\begin{equation}
\tau_i|0\rangle^{\otimes N}=|0\rangle^{\otimes i-1}\otimes \left[\frac{1}{\sqrt{2}}\left(|00\rangle+|11\rangle\right)\right]_{i,i+1}\otimes|0\rangle^{\otimes N-i-1}.
\end{equation}

Now we turn the Pauli version of braid representation into the Majorana version. Majorana operators are well connected to Pauli matrices under Jordan-Wigner transformation, 
\begin{equation}\label{MF}
\gamma_{2j-1}=\left[\prod_{k=1}^{j-1}\sigma_k^z\right]\sigma_j^x, \quad \gamma_{2j}=\left[\prod_{k=1}^{j-1}\sigma_k^z\right]\sigma_j^y.
\end{equation}
We see from above definition that one spin site corresponds to two Majorana sites. The Majorana operators are Hermitian and satisfy Clifford algebra,
\begin{equation}
\gamma_i=\gamma_i^{\dag}, \quad \{\gamma_i,\, \gamma_j\}=2\delta_{ij}.
\end{equation}
Substituting Eq. (\ref{MF}) into Eq. (\ref{BraidPauli}), one obtains the Majorana representation of braid generators, as
\begin{equation}\label{BraidMaj}
\tau_i=\frac{1}{\sqrt{2}}\left(1+\gamma_{2i-1}\gamma_{2i+1}\right)=e^{\tfrac{\pi}{4}\gamma_{2i-1}\gamma_{2i+1}}.
\end{equation}
This braid representation, as was presented in many papers, describes the non-Abelian statistics properties of Majorana zero modes\cite{kitaev2001unpaired,ivanov2001non,alicea2011non}. However, different from braiding nearest Majorana zero modes as it usually appears in papers, the braid generators that we define in Eq. (\ref{BraidMaj}) include only odd number Majorana sites and transform the odd Majorana operators into
\begin{equation}\label{PM}
\tau_i\gamma_{2j-1}\tau_i^{\dag}=\left\{
\begin{aligned}
&\gamma_{2j-1}, \quad &&\text{if}\,\,j\notin\{i,\,i+1\},\\
&-\gamma_{2i+1}, \quad &&\text{if}\,\, j=i,\\
&\gamma_{2i-1}, \quad &&\text{if}\,\, j=i+1.
\end{aligned}
\right.
\end{equation}
We find that the operations of braid generators on Majorana operators are equivalent to exchanging two odd-nearest  Majorana operators(up to a sign).  Let us define the braiding operation $B_{p,q}$ exchanging only two Majorana operators $\gamma_p$ and $\gamma_q$ with odd $p$ and $q$, 
\begin{equation}
B_{p,q}\gamma_p B_{p,q}^{\dag}\propto\gamma_q, \quad B_{p,q}\gamma_q B_{p,q}^{\dag}\propto\gamma_p, 
\end{equation}
while all other Majorana operators are not exchanged except $p$ and $q$. Here symbol ``$\propto$'' means that the result is up to a sign. $B_{p,q}$ represents a set of different braid operators sharing the same property: In diagrammatic version of the set of braid operators $B_{p,q}$, the $p$-th strand ends at the $q$-th strand site, and the $q$-th strand ends at the $p$-th strand site, regardless of the concrete path of the strand.

In this paper, we choose $|0\rangle^{\otimes N}$ as the initial N-qubit state. Indeed, from the view point of stabilizer code, the state is stabilized by a set of N independent operators with all eigenvalues -1,
\begin{equation}
\begin{aligned}
\mathcal{S}&=\{\sigma_1^z,\, \sigma_2^z,\, \cdots \, \sigma_{N-1}^z, \,\sigma_N^z\}\\
&=\{i\gamma_2\gamma_1,\, i\gamma_4\gamma_3,\, \cdots \, i\gamma_{2N-2}\gamma_{2N-3},\, i\gamma_{2N}\gamma_{2N-1}\}.
\end{aligned}
\end{equation}
For the N-qubit space, if there are N independent stabilizers, then the logical space should have dimension $2^{N-N}=1$, i.e., there is only one common eigenstate $|0\rangle^{\otimes N}$ for all stabilizers in $\mathcal{S}$ with the same eigenvalue -1. 

Applying any braid operator $\mathcal{B}$ to the initial state $|0\rangle^{\otimes N}$ is equivalent to changing the stabilizer set $\mathcal{S}$ into (up to a sign)
\begin{equation}
\begin{aligned}
\mathcal{S}'&=\mathcal{B}\mathcal{S} \mathcal{B}^\dag\\
&\sim\{i\gamma_2\gamma_{p(1)},\, i\gamma_4\gamma_{p(3)},\, \cdots \, i\gamma_{2N}\gamma_{p(2N-1)} \},
\end{aligned}
\end{equation}
where $p$ stands for a permutation of the odd site number of the Majorana operators. The final state $|\Psi\rangle=\mathcal{B}|0\rangle^{\otimes N}$ is stabilized by $\mathcal{S}'$ with all eigenvalues -1. In this paper, we will use the final stabilizer set $\mathcal{S}'$ to classify the entanglement.  As was shown in Eq. (\ref{PM}), for the stabilizer set , the sign before the Majorana operators is not important and can be ignored. Indeed, for the Majorana representation, the permutation group $S_N$ is the quotient group of braid group $B_N$. Hence the only useful part of the braid group  in our paper is the permutation group $S_N$, which describes the permutation of odd number Majorana sites.

\section{Some simple examples of Generating entanglement by braiding}\label{Sec3}
In this section, to give an intuitive description about the relationship between braids and entanglement, we consider some simple cases.
\subsection{Permuting two Majorana operators}
Let us consider the simplest example about relationships between braiding and entanglement  by exchanging only two Majorana operators $\gamma_{2a-1}$ and $\gamma_{2b-1}$ that correspond to spin sites $a$ and $b$ respectively.
We denote the braiding exchange of $\gamma_{2a-1}$ and $\gamma_{2b-1}$ by cyclic permutation $P=(ab)$. Then the final stabilizer set $\mathcal{S}'$ is (up to a sign)
\begin{equation}
\mathcal{S}'=\{i\gamma_2\gamma_1,\,  \cdots \, i\gamma_{2a}\gamma_{2b-1},\,\cdots \, i\gamma_{2b}\gamma_{2a-1},\, \cdots \,  i\gamma_{2N}\gamma_{2N-1} \}.
\end{equation}
Except $a$ and $b$, all other spin sites in the final state $|\Psi\rangle$ must be separable with each others since only two stabilizers $i\gamma_{2a}\gamma_{2b-1}$ and $i\gamma_{2b}\gamma_{2a-1}$ in $\mathcal{S}'$ are different from the initial $\mathcal{S}$ and the stabilized space is $2^{N-N}=1$ dimensional. In other words,  we only need to check whether the $a$-th and $b$-th spin sites are entangled or not in the final state $|\Psi\rangle$.

Now we prove that the $a$-th and $b$-th sites must be entangled. We denote the 2-qubit state on $a$ and $b$ sites by $|\varphi\rangle_{ab}$. Substituting Eq. (\ref{MF}) into two stabilizers $i\gamma_{2a}\gamma_{2b-1}$ and $i\gamma_{2b}\gamma_{2a-1}$, one obtains (suppose $a<b$)
\begin{eqnarray}
&i\gamma_{2a}\gamma_{2b-1}=-\mathbb{I}^{\otimes a-1}\otimes\sigma_{a}^y\otimes[\sigma^z]^{\otimes b-a-1}\otimes \sigma_b^y\otimes \mathbb{I}^{\otimes N-b},\quad \\
&i\gamma_{2b}\gamma_{2a-1}=-\mathbb{I}^{\otimes a-1}\otimes\sigma_{a}^x\otimes[\sigma^z]^{\otimes b-a-1}\otimes \sigma_b^x\otimes \mathbb{I}^{\otimes N-b}.\quad
\end{eqnarray}
Obviously, $|\varphi\rangle_{ab}$ must be common eigenstate of $\sigma_a^y\otimes\sigma_b^y$ and $\sigma_a^x\otimes\sigma_b^x$ due to the stabilizer condition. If $|\varphi\rangle_{ab}$ is not entangled, say, $|\varphi\rangle_{ab}=|\phi\rangle_a\otimes|\phi\rangle_b$, $|\phi\rangle_a$ must be common eigenstate of $\sigma_a^y$ and $\sigma_a^x$. But $\sigma_a^y$ and $\sigma_a^x$ cannot share the same eigenstate. Hence, $|\varphi\rangle_{ab}$ is an entangled 2-qubit state. Hence we conclude that only if strand $a$ in braid diagram ends at the position of strand $b$, the corresponding $a$ and $b$-th spin sites are entangled in the final state $|\Psi\rangle$.

\subsection{Permuting Three Majorana operators}
Now we discuss the case that three Majorana operators $\gamma_{2a-1}$, $\gamma_{2b-1}$ and $\gamma_{2c-1}$ are permuted under the braiding operation. Similar to the previous section, we denote the three Majorana operators' permutation by cyclic notation $P=(abc)$, which describes strand $a$ ends at strand position $b$, strand $b$ ends at position $c$, strand $c$ ends at position $a$ in braid diagram. After the braiding operation, the stabilizer set of final state $|\Psi\rangle$ becomes (up to a sign)
\begin{equation}
\mathcal{S}'\sim\{i\gamma_{2}\gamma_{1},...\,i\gamma_{2a}\gamma_{2b-1},...\,i\gamma_{2b}\gamma_{2c-1},...\, i\gamma_{2c}\gamma_{2a-1},...\}.
\end{equation}
In comparison with the initial stabilizer set $\mathcal{S}$, $\mathcal{S}'$ only has three different stabilizers $\{i\gamma_{2a}\gamma_{2b-1},\, i\gamma_{2b}\gamma_{2c-1},\, i\gamma_{2c}\gamma_{2a-1}\}$. Due to the stabilizer condition, spin sites other than $a$, $b$ and $c$ must be separable with each others. Now let us focus on the 3-qubit subsystem final state $|\varphi\rangle_{abc}$ stabilized by $\{i\gamma_{2a}\gamma_{2b-1},\,i\gamma_{2b}\gamma_{2c-1},\, i\gamma_{2c}\gamma_{2a-1}\}$ on sites $a$, $b$ and $c$.   We will prove that $|\varphi\rangle_{abc}$ cannot be separated into any two parties. For 3-qubit case, we only need to prove that any one qubit is entangled with another two. Without loss of generality, if we suppose $|\varphi\rangle_{abc}=|\phi_1\rangle_a\otimes|\phi_2\rangle_{bc}$,  the $a$-site part of all stabilizers must commute. But it is easy to find two stabilizers $i\gamma_{2a}\gamma_{2b-1}$ and $i\gamma_{2a-1}\gamma_{2a}\gamma_{2b-1}\gamma_{2b}\gamma_{2c-1}\gamma_{2c}$, so that the $a$-site subsystem operators in them do not commute. Concretely, in Pauli version, the $a$-site part of $i\gamma_{2a}\gamma_{2b-1}$ is $\sigma_a^{x(y)}$(here $x$ or $y$ only depends on which one of $a$ and $b$ is larger), while the $a$-site part of $i\gamma_{2a-1}\gamma_{2a}\gamma_{2b-1}\gamma_{2b}\gamma_{2c-1}\gamma_{2c}$ is $\sigma_a^z$. Clearly, it violates our assumption and means that $|\varphi\rangle_{abc}$ cannot be separated into $a$ and $bc$ parties. Similar constructions can also be applied to $b-ac$ and $c-ab$ cases. Hence, three sites $a$, $b$ and $c$ are entangled in the final state $|\Psi\rangle$.

\subsection{Permuting Four Majorana operators}
Now we consider the braiding operation that permutes four Majorana operators with only 1 sub-cyclic permutation $P=(abcd)$. Then only four stabilizers  $\{i\gamma_{2a}\gamma_{2a-1},\,i\gamma_{2b}\gamma_{2b-1},\, i\gamma_{2c}\gamma_{2c-1},\, i\gamma_{2d}\gamma_{2d-1}\}$ in initial stabilizer set $\mathcal{S}$ are changed by braiding operation nontrivially into $\{i\gamma_{2a}\gamma_{2b-1},\, i\gamma_{2b}\gamma_{2c-1},\, i\gamma_{2c}\gamma_{2d-1},\, i\gamma_{2d}\gamma_{2a-1}\}$ in final stabilizer set $\mathcal{S}'$. Since qubit sites other than $a,b,c \,\text{and}\,d$ must be separable, now we prove that the final 4-qubit state $|\varphi\rangle_{abcd}$ on $a,b,c,d$ sites are entangled, i.e., the 4-qubit state cannot be separated into any 2 parties. There are two cases to be proved. The first case is the entanglement between 1-qubit and 3-qubit, and the second case is the entanglement between 2-qubit and 2-qubit. 

Let us consider the first case. To prove $a$-site and $bcd$-site are not separable, one only needs to find two stabilizers from $\mathcal{S}'$ so that their corresponding $a$-site parts do not commute. We can choose $i\gamma_{2a}\gamma_{2b-1}$ with $a$-site part $\sigma_{a}^{x}$ or $\sigma_{a}^{y}$ and $\gamma_{2a-1}\gamma_{2a}\gamma_{2b-1}\gamma_{2b}\gamma_{2c-1}\gamma_{2c}\gamma_{2d-1}\gamma_{2d}$ with $a$-site part $\sigma_{a}^z$. Since $[\sigma_{z}^{x(y)},\, \sigma_a^z]\neq0$, $a$-site and $bcd$-site must be not separable. Similar results can also be applied to  $b-acd$, $c-abd$ and $d-abc$ cases. 

Now we consider the second case. Our goal is still finding non-commuting subsystem operators from stabilizers in $\mathcal{S}'$. We first choose one stabilizer operator 
\begin{equation}
\begin{aligned}
\gamma_{2a-1}\gamma_{2a}\gamma_{2b-1}\gamma_{2b}\gamma_{2c-1}\gamma_{2c}\gamma_{2d-1}\gamma_{2d}\propto \sigma_a^z\otimes\sigma_b^z\otimes\sigma_c^z\otimes\sigma_d^z.
\end{aligned}
\end{equation}
In $\{i\gamma_{2a}\gamma_{2b-1},\,i\gamma_{2b}\gamma_{2c-1},\, i\gamma_{2c}\gamma_{2d-1},\, i\gamma_{2d}\gamma_{2a-1}\}$, each stabilizer has only 2 spin sites that are not $\sigma^{z}$ or $\mathbb{I}$. For example, for $i\gamma_{2a}\gamma_{2b-1}$ in Pauli version, only the operators on sites $a$ and $b$ are $\sigma^x$ or $\sigma^y$, while all of the other sites are $\sigma^{z}$ or $\mathbb{I}$.  If we want to prove that $ab$-site and $cd$-site are entangled, we can choose $i\gamma_{2d}\gamma_{2a-1}$ with $ab$-site part $\sigma_{a}^{x(y)}\otimes\sigma_{b}^{z}$ or $\sigma_{a}^{x(y)}\otimes\mathbb{I}_b$ together with $\gamma_{2a-1}\gamma_{2a}\gamma_{2b-1}\gamma_{2b}\gamma_{2c-1}\gamma_{2c}\gamma_{2d-1}\gamma_{2d}$ with $ab$-site part $\sigma_{a}^{z}\otimes\sigma_{b}^{z}$. Then we find two non-commuting operators on $ab$-sites from stabilizer sets. Hence $ab$-site and $cd$-site must be entangled. Similar proof can be also applied to $ac-bd$ and $ad-bc$ cases. 

In summary, the 4-qubit sites $a$, $b$, $c$ and $d$ are not separable in the final state $|\Psi\rangle$.


\section{Braiding and entanglement for multiqubits}\label{Sec4}

In the previous section, we give some simple examples about the relationship between  braids and few qubits entanglement. Now we extend  the cases to multi-qubit system. We consider two special types of permutation. The first type includes only one sub-cyclic permutation, and the second type includes two sub-cyclic permutations. 

\subsection{Permuting (r+s) Majorana operators with $P=(a_1a_2...a_r...a_{r+s})$}
Now we consider the braiding operation permuting (r+s) Majorana operators with only one sub-cyclic permutation $P=(a_1a_2...a_r...a_{r+s})$. Here $r$ and $s$ are arbitrary positive integers satisfying $r+s\leq N$, and $\{a_i\,|\, i\in[1,\,r+s]\}$ represent arbitrary different $(r+s)$ spin sites. As was discussed in previous sections, the spin sites not belonging to $\{a_i\,|\, i\in[1,\,r+s]\}$ must be still separate with each others due to the stabilizer condition.  Let $|\varphi\rangle_{a_1...a_{(r+s)}}$ be the subsystem state of the final state $|\Psi\rangle$. Now we prove that $|\varphi\rangle_{a_1...a_{(r+s)}}$ on sites $\{a_i\,|\, i\in[1,\,r+s]\}$ is an entangled $(r+s)$-qubit state, i.e., the state cannot be separated into any two parties. 

Let us consider two parties: one party includes sites $\{b_{j}\,|\, j\in[1,r]\}$, the other party includes sites  $\{b_{j}\,|\, j\in[r+1,r+s]\}$, where $\{b_{j}\,|\, j\in[1,r+s]\}=\{a_{i}\,|\, i\in[1,r+s]\}$. Here we choose the new notation $\{b_{j}\,|\, j\in[1,r+s]\}$ instead of  $\{a_{i}\,|\, i\in[1,r+s]\}$ to ensure that the entangled parties are irrelevant to the permutation order. Due to the permuting operation $P=(a_1a_2...a_r...a_{r+s})$, there must be at least one Majorana operator $\gamma_{2b_{m}-1}$($m\in[1,r]$) that is permuted into $\gamma_{2b_{n}-1}$($n\in[r+1,r+s]$) by the braiding operation. In other words, $i\gamma_{2b_{m}}\gamma_{2b_{n}-1}$ must be a stabilizer of the final state $|\Psi\rangle$. Then the  $\{b_{j}\,|\, j\in[1,r]\}$ party of the Pauli version of  $i\gamma_{2b_{m}}\gamma_{2b_{n}-1}$ can be expressed as
\begin{equation}
\Gamma_1=(\sigma_{b_1}^z)^{c_1}\otimes(\sigma_{b_2}^z)^{c_2}\otimes\cdots \otimes\sigma_{b_m}^{x(y)}\otimes\cdots\otimes(\sigma_{b_r}^z)^{c_r},
\end{equation}
where each $c_i$ corresponds to the power of the operator on site $b_i$, and $\{c_1,\, ...c_{m-1},\, c_{m+1},...c_{r}\}=0\,\text{or} \,1$.
Another stabilizer we need is $\prod_{i=1}^{r+s}(\gamma_{2b_{i}-1}\gamma_{2b_i})$, whose $\{b_{j}\,|\, j\in[1,r]\}$-site party is 
\begin{equation}
\Gamma_2=\sigma_{b_1}^z\otimes\sigma_{b_2}^z\otimes\cdots\otimes\sigma_{b_m}^z\otimes\cdots\otimes\sigma_{b_r}^z.
\end{equation}
It is easy to check that $[\Gamma_1\, \Gamma_2]\neq0$, then the $\{b_{j}\,|\, j\in[1,r]\}$ party and $\{b_{j}\,|\, j\in[r+1,r+s]\}$ party must be entangled. Here $r$ and $s$ can be any positive integers satisfying $r + s \leq N$, hence $|\varphi\rangle_{a_1...a_{(r+s)}}$ is an entangled subsystem state of the final state $|\Psi\rangle$. 

\subsection{Permuting (r+s) Majorana operators with $P=(a_1a_2 \cdots a_r)(b_{1}b_{2} \cdots b_{s})$}
Now we consider the braiding operation that permutes Majorana operators in the case with two sub-cyclic permutations $P=(a_1a_2 \cdots a_r)(b_{1}b_{2} \cdots b_{s})$, here $a_i$ and $b_i$ are irrelevant to the notations in previous sections. Since the final state corresponding to permutation process $P=(a_1a_2\cdots a_r)$ has been proved to be entangled, here we only need to consider whether the two parties $\{a_i|i\in[1,r]\}$-site and $\{b_i|i\in[1,s]\}$-site are entangled or not. After braiding operation on the initial stabilizer set, the changed stabilizers on $\{a_i|i\in[1,r]\}$ and $\{b_i|i\in[1,s]\}$ parties are $\mathcal{S}'_a=\{i\gamma_{2a_1}\gamma_{2a_2-1},\,i\gamma_{2a_2}\gamma_{2a_3-1},\,...\,i\gamma_{2a_r}\gamma_{2a_1-1}\}$ and $\mathcal{S}'_b=\{i\gamma_{2b_1}\gamma_{2b_2-1},\,i\gamma_{2b_2}\gamma_{2b_3-1},\,...\,i\gamma_{2b_s}\gamma_{2b_1-1}\}$ respectively. In the following discussion, we only need to consider the changed stabilizers after braiding operation because the unchanged stabilizers act trivially on the sites $\{a_i|i\in[1,r]\}$ and $\{b_i|i\in[1,s]\}$.  There are totally three cases to be discussed.

\begin{enumerate}
\item $\max\{a_i|i\in[1,r]\}<\min\{b_i|i\in[1,s]\}$.

In this case,  due to the condition $\max\{a_i|i\in[1,r]\}<\min\{b_i|i\in[1,s]\}$, it is obvious that all the $\{a_i|i\in[1,r]\}$ party of stabilizers in $\mathcal{S}'$ commute with each others. Hence, the $\{a_i|i\in[1,r]\}$ and $\{b_i|i\in[1,s]\}$ parties are separable in the final state.

\item\label{C2} $\min\{a_i|i\in[1,r]\}<\min\{b_i|i\in[1,s]\}<\max\{a_i|i\in[1,r]\}<\max\{b_i|i\in[1,s]\}$.

In this case, we prove that after braiding operation, the $\{a_i|i\in[1,r]\}$ and $\{b_i|i\in[1,s]\}$ parties are entangled. We denote the permutation processing $P=(a_1a_2 \cdots a_r)(b_{1}b_{2} \cdots b_{s})$ by $P=P_a\cdot P_b$, where $P_a=(a_1a_2 \cdots a_r)$ and $P_b=(b_{1}b_{2} \cdots b_{s})$. In combination with the condition $\min\{a_i|i\in[1,r]\}<\min\{b_i|i\in[1,s]\}<\max\{a_i|i\in[1,r]\}<\max\{b_i|i\in[1,s]\}$, to preserve the permutation $P_a$ and $P_b$, there must be two stabilizers $i\gamma_{2a_p}\gamma_{2a_q-1}$ and $i\gamma_{2b_j}\gamma_{2b_k-1}$ in $\mathcal{S}'$ so that 
$a_p<b_j<a_q<b_k$, where $\{a_p,a_q\}\subseteq\{a_i|i\in[1,r]\}$, $\{b_j,b_k\}\subseteq\{b_i|i\in[1,s]\}$,
\begin{eqnarray}
&&\begin{split}&i\gamma_{2a_p}\gamma_{2a_q-1} \\
&\propto \mathbb{I}^{\otimes a_p-1}\otimes \sigma_{a_p}^{x}\otimes (\sigma^{z})^{\otimes a_q-a_p-1}\otimes\sigma_{a_q}^{x}\otimes (\mathbb{I})^{\otimes N-a_q},\end{split}\label{G1}\\
&&\begin{split}&i\gamma_{2b_j}\gamma_{2b_k-1}\\
&\propto \mathbb{I}^{\otimes b_j-1}\otimes \sigma_{b_j}^{x}\otimes (\sigma^{z})^{\otimes b_k-b_j-1}\otimes\sigma_{b_k}^{x}\otimes (\mathbb{I})^{\otimes N-b_k}.\end{split}\label{G2}
\end{eqnarray}
The $\{a_i|i\in[1,r]\}$ parties of Eq. (\ref{G1}) and Eq. (\ref{G2}) are
\begin{eqnarray}
&&\begin{split}&i\gamma_{2a_p}\gamma_{2a_q-1} \longrightarrow \Gamma_3:\\
&(\sigma_{a_1}^z)^{u_1}\otimes(\sigma_{a_2}^z)^{u_2}...\otimes \sigma_{a_p}^{x}\otimes...\otimes\sigma_{a_q}^{x}\otimes...\otimes (\sigma_{a_r}^z)^{u_{r}},\end{split}\label{G1a}\\
&&\begin{split}&i\gamma_{2b_j}\gamma_{2b_k-1}\longrightarrow  \Gamma_4:\\
&(\sigma_{a_1}^z)^{v_1}\otimes(\sigma_{a_2}^z)^{v_2}...\otimes \mathbb{I}_{a_p}\otimes...\otimes\sigma_{a_q}^{z}\otimes...\otimes (\sigma_{a_r}^z)^{v_{r}},\end{split}\label{G2a}
\end{eqnarray}
where $u_i$ and $v_i$ correspond to the power of operators on site $a_i$, and  $\{u_1,u_2...u_{p-1}, u_{p+1},... u_{q-1}, u_{q+1},...u_r\}=0 \,\text{or}\, 1$, $\{v_1,v_2...v_{q-1}, v_{q+1},...v_r\}=0 \,\text{or}\, 1$.  We can see from Eq. (\ref{G1a}) and Eq. (\ref{G2a}) that only the $a_q$-site parties of $\Gamma_3$ and $\Gamma_4$ do not commute, hence   $[\Gamma_3,\Gamma_4]\neq0$. Hence, the $\{a_i|i\in[1,r]\}$ and $\{b_i|i\in[1,s]\}$ parties are entangled in the final state $|\Psi\rangle$.

\item $\min\{a_i|i\in[1,r]\}<\min\{b_i|i\in[1,s]\}<\max\{b_i|i\in[1,s]\}<\max\{a_i|i\in[1,r]\}$.
\begin{enumerate}
\item $\forall a_j\notin\left[\min\{b_i|i\in[1,s]\},\, \max\{b_i|i\in[1,s]\}\right],\, j\in[1,r]$. 

In this case, we prove that the $\{a_i|i\in[1,r]\}$ party and $\{b_i|i\in[1,s]\}$ parties are separable.  Let us first consider the stabilizers in set $\mathcal{S}'_a$. It is known that the stabilizers commute with each others, and all of the sites other than $\{a_i|i\in[1,r]\}$ of the stabilizers in $\mathcal{S}'_a$ must be $\sigma^z$ or $\mathbb{I}$. Hence, the $\{a_i|i\in[1,r]\}$ parties of all stabilizers in $\mathcal{S}'_a$ must commute. Secondly, for the stabilizers in set $\mathcal{S}'_b$, due to the condition $\forall a_j\notin\left[\min\{b_i|i\in[1,s]\},\, \max\{b_i|i\in[1,s]\}\right], \, j\in[1,r]$, the  $\{a_i|i\in[1,r]\}$ parties of the stabilizers in set $\mathcal{S}'_b$  must be identity. Then all of the $\{a_i|i\in[1,r]\}$ parties of the stabilizers in set $\mathcal{S}'$ commute with each others. Hence in this case, the $\{a_i|i\in[1,r]\}$ party and $\{b_i|i\in[1,s]\}$ parties are separable in the final state $|\Psi\rangle$.
\item $\exists\, a_j\in\left[\min\{b_i|i\in[1,s]\},\, \max\{b_i|i\in[1,s]\}\right],\, j\in[1,r]$. 

This case is similar to the Case \ref{C2} we mentioned above. To preserve the permutation processing $P=P_a\cdot P_b$. we can always find at least two stabilizers $i\gamma_{2a_p}\gamma_{2a_q-1}$ and $i\gamma_{2b_j}\gamma_{2b_k-1}$ in $\mathcal{S}'$ so that 
$a_p<b_j<a_q<b_k$ or $b_j<a_p<b_k<a_q$, where $\{a_p,a_q\}\subseteq\{a_i|i\in[1,r]\}$, $\{b_j,b_k\}\subseteq\{b_i|i\in[1,s]\}$. Then the result of Case \ref{C2} can be applied here directly. 
\end{enumerate}

%

\end{enumerate}

\section{Conclusion and Discussion}
In summary, by analyzing the properties of the final stabilizer set after braiding operations, we obtain the entanglement properties of the final stabilized state $|\Psi\rangle$. Our proof ends at the case including only two sub-cyclic permutations. However, braiding operations permuting Majorana operators under the multi sub-cyclic permutations $P=(a_1a_2...a_r)(b_1b_2...b_s)(c_1...c_t)\cdots (d_1...d_u)$ can be discussed in a similar way like the  two sub-cyclic case $P=(a_1a_2...a_r)(b_1b_2...b_s)$. Here we recall that an entangled state in our paper is defined by the non-separability of the state into any two parties. To check whether two parties are entangled or not, one only needs to find two non-commuting operators  on the sites of one party from the final stabilizer set $\mathcal{S}'$. If there exist two non-commuting operators, then the two parties are entangled; If not, then the two parties must be separable due to the dimension of the stabilized space is only $2^{N-N}=1$.

Our results show the close relationships between quantum entanglement and the permutation of the strands in the diagrammatic version under braiding operations.   The results rely on the Majorana fermionic representation of braids. Further extension of the results may be related to the $\mathbb{Z}_3$ parafermionic representation of braids \cite{fendley2012parafermionic,yu2016z3}, which is also related to the local unitary representation of the braids\cite{Rowell2012}.

\section*{Acknowledgments}
The author would like to thank Professor Z. Wang for his helpful discussions and encouragements. This work is in part supported by NSF of China (Grant No. 11475088) and China Scholarship Council(CSC). 

\bibliographystyle{apsrev4-1.bst}
%
\end{document}